# Warm absorbers in active galactic nuclei

C. S. Reynolds and A. C. Fabian
*Institute of Astronomy, Madingley Road, Cambridge CB3 0HA*

**ABSTRACT**
Recent *ASCA* observations confirm the presence of X-ray absorption due to partially ionized gas in many Seyfert 1 galaxies; the so-called warm absorber. Constraints on the location of the warm material are presented with the conclusion that this material lies at radii coincident with, or just outside, the broad-line region. The stability of this warm material to isobaric perturbations under the assumptions of thermal and photoionization equilibrium is also studied. It is shown that there is a remarkably small range of ionization parameter, $\xi$, for which the warm absorber state is stable. The robustness of this result to changes in the shape of the primary continuum, the assumed density and optical depth is investigated. Given the constraints on the location and the stability properties of the material, several models for the environments of Seyfert nuclei are discussed. These attempt to explain the presence of significant amounts of partially ionized material. In particular, various models of the broad-line region are discussed. The simple two-phase model of the broad-line region proves to be unsatisfactory. A model of the broad-line region is presented in which a turbulent, hot intercloud medium is mechanically heated. Turbulent mixing layers could then give rise to warm absorption features. Finally, a model is discussed in which the warm absorber is due to a steady state, radiatively driven outflow.

## 1 INTRODUCTION

X-ray reprocessing in active galactic nuclei (AGN) significantly affects the observed X-ray spectrum. A study of this reprocessing is important since it probes the geometry and physical state of matter in the central regions of AGN. The reflection of X-rays from optically thick cold material (e.g. an accretion disc) is invoked to explain the spectra of Seyfert 1 galaxies which often display a 6.4-keV fluorescent K-line of cold iron and a broad bump peaking at $\sim 20\,\mathrm{keV}$ (Guilbert & Rees 1988; Lightman & White 1988; George & Fabian 1991; Matt, Perola & Piro 1991). During the past decade it has been realized that partially ionized, optically thin gas along our line of sight to the central X-ray source can also have a dramatic effect on the soft X-ray spectrum. This partially ionized material has become known as the warm absorber. The observational signatures of warm absorbers are discussed on a theoretical basis by Netzer (1993).

The presence of such gas was initially postulated to explain the unusual shape of the absorption needed to fit the soft and hard X-ray spectrum of the QSO MR 2251−178 (Halpern 1984; Pan, Stewart & Pounds 1990) and as a component in Seyfert 2 galaxies (Krolik & Kallman 1987). Further evidence came from *Ginga* which found deeper iron K-edges in some Seyfert 1 galaxies than predicted by the simple reflection model (Nandra, Pounds & Stewart 1990; Nandra et al. 1991; Nandra & Pounds 1994). *ROSAT* position sensitive proportional counter (PSPC) observations of some Seyfert 1 galaxies also suggested the presence of absorption K-edges due to O VII and O VIII (Nandra & Pounds 1992; Nandra et al. 1993; Fiore et al. 1993; Turner et al. 1993). However, other explanations of the data (such as a multiple-component primary source or partial covering by cold absorbing material) could not firmly be ruled out by these data.

The superior spectral energy resolution of the *ASCA* solid-state imaging spectrometer (SIS) allows, for the first time, X-ray spectral features to be accurately identified and measured. This instrument (unlike any previous instrument) is capable of separating the O VII and O VIII K-edges (at rest energies of $0.74\,\mathrm{keV}$ and $0.87\,\mathrm{keV}$ respectively). Fabian et al. (1994) have studied the bright Seyfert 1 galaxy MCG−6−30−15 ($z=0.008$) with *ASCA*. They find clear evidence for O VII and O VIII edges in the X-ray spectrum, thereby confirming the presence of a warm absorber. They use the photoionization code CLOUDY (Ferland 1991) to construct a grid of one-zone models in which the primary continuum photoionizes a geometrically thin shell of gas. These models are characterized by a column density, $N_\mathrm{W}$, and an ionization parameter, $\xi$, defined by

$$\xi \equiv \frac{L}{nR^2}. \tag{1}$$

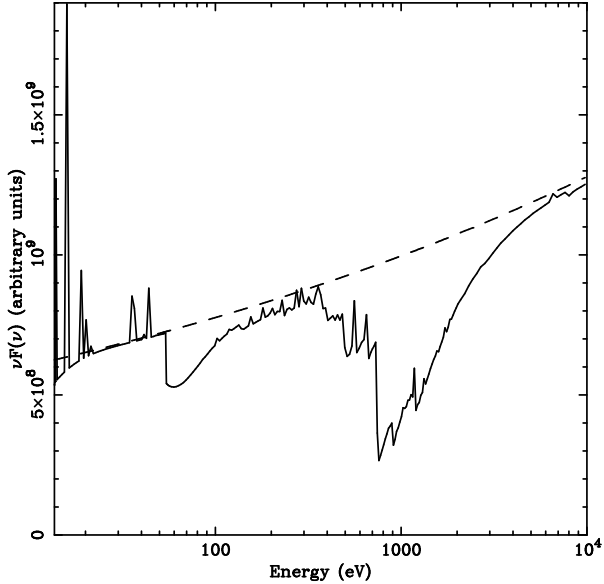

**Figure 1.** Theoretical spectrum produced by the passage of a primary powerlaw spectrum through a thin shell of warm absorbing gas (solid line). The parameters of the model are taken to be those derived from recent *ASCA* observations of MCG−6−30−15: i.e. the photon index is $\Gamma = 1.9$, the ionization parameter of the warm material is $\xi = 35 \,\mathrm{erg\,cm\,s^{-1}}$ and the warm column density is $N_\mathrm{W} = 10^{22} \,\mathrm{cm}^{-2}$. A radius of $10^{16}$ cm is assumed. The effects of emission and reflection from the warm material (assuming unit covering fraction) as well as absorption are included. The dashed line represents the incident powerlaw spectrum.

When such models are fitted to a 30000-s *ASCA* SIS observation of MCG−6−30−15, the best-fitting model parameters are $N_\mathrm{W} = 10^{22\pm 0.02}\,\mathrm{cm}^{-2}$ and $\xi = 35 \pm 2 \,\mathrm{erg\,cm\,s^{-1}}$. This corresponds to gas at $10^5$ K. Fig. 1 shows the theoretical absorbed spectrum (from 13.6 eV to 10 keV) for a warm absorber with these parameters. This includes the effects of emission and reflection (assuming a unit covering fraction) as well as absorption. Strong He II, O VII and O VIII edges are seen at 54.4 eV, 0.74 keV and 0.87 keV respectively.

Fabian et al. (1994) also find variability in the warm absorber between two *ASCA* observations separated by 3 weeks. During this 3-week period, the ionizing flux decreases while both the ionization parameter and column density increase. Section 3 discusses this in more detail.

Warm absorption features are common in the X-ray spectra of Seyfert 1 galaxies and narrow emission line galaxies (NELGs). In the *Ginga* sample of Nandra & Pounds (1994), 9 out of 20 Seyfert 1 galaxies show evidence for warm absorbers, as do 3 out of 7 NELGs. Therefore, in these sources, the warm absorbing material must have a large covering fraction as seen from the central source (perhaps ∼ 50 per cent, at least along lines of sight accessible to us). The absorption features seen in BL Lac objects (Madejski et al. 1991) may also be due to a warm absorber (possibly entrained in a jet). However, warm absorbers seem to be absent in other radio-loud quasars and powerful radio-quiet quasars. An exception to this is the warm X-ray/UV absorber in 3C351 (Mathur et al. 1994).

The present paper discusses the warm absorbing material in the context of the standard models for Seyfert galaxies and other AGN. Section 2 discusses the physical state and emission line spectrum of the warm absorbing material. Section 3 attempts to place constraints on the location of the warm absorber and concludes that this material exists at radii characteristic of the broad-line region (BLR). Section 4 examines the stability of the warm absorber to isobaric perturbations. It is found that the warm absorbing state is only stable for a remarkably small range of ionization parameter $\xi$ unless there is a very soft primary spectrum. Section 5 presents possible physical models for the warm absorber. In particular, models relating this gas to the BLR are assessed. A new BLR model is presented in which a turbulent hot intercloud medium is mechanically heated by the action of radiation pressure on the broad-line clouds. Warm absorption features might be expected from the turbulent mixing layers formed within such a scenario. Section 6 provides a summary of the issues raised.

## 2 PHYSICAL STATE OF THE WARM ABSORBER

CLOUDY predicts the detailed physical state of the warm material, given the assumption of photoionization and thermal equilibrium.

We shall take an ionization parameter of $\xi = 30\,\mathrm{erg\,cm\,s^{-1}}$ and a column density of $N_\mathrm{W} = 10^{22}\,\mathrm{cm}^{-2}$ (as inferred from *ASCA* observations of MCG−6−30−15). Given these parameters the temperature of the material is $10^5$ K. The dominant source of heating is photoelectric absorption by metals (85 per cent of the total heating). Photoelectric absorption by helium accounts for much of the remaining heating (11 per cent of the total heating). Cooling is dominated by the collisionally excited lines of O VI $\lambda 1035$ (30 per cent of the total cooling) and Ne VIII $\lambda 774$ (20 per cent of the total cooling). For unit covering fraction the luminosity in these lines can be as much as a few per cent of the total ionizing luminosity of the central engine. Other significant UV/optical lines predicted are Ly$\alpha$, C IV $\lambda 1549$ and Fe XIV $\lambda 5303$.

Interestingly, *IUE* observations of MCG−6−30−15 show that the high-ionization lines are very weak. Absorption by neutral material has been invoked to explain the weak high-ionization lines. However, X-ray observations suggest that no cold absorption (extra to Galactic absorption) is present. There are several possible explanations for this apparent discrepancy. First, the geometry could be such that the material emitting the high-ionization lines is obscured whilst the X-ray emitting region is unobscured. The required geometry would be impossible to realize within the standard model of Seyfert galaxies. Secondly, a dusty warm absorber or outer BLR could lead to significant optical/UV extinction without affecting the X-rays. Such a model predicts near-infrared thermal emission from the hot dust. Finally, MCG−6−30−15 may have an intrinsically unusual BLR in which high-ionization lines are not produced by broad-line clouds and the warm absorber is restricted to our line of sight. The emission of, say, C IV from the warm absorber may therefore be much weaker that that predicted assuming a large covering fraction. Such a model is problematic given the frequency with which warm absorbers are observed.

We note in passing that the high-ionization lines from the warm absorber may severely complicate and confuse conventional reverberation mapping of the BLR.

## 3 LOCATION OF THE WARM ABSORBER

Given some reasonable assumptions, simple arguments can be used to constrain the distance of the warm absorbing gas from the central engine. In order to derive numerical constraints, we shall use parameters appropriate to MCG−6−30−15. Thus, we shall suppose that the warm material has a measured ionization parameter of $\xi = 30\,{\rm erg\,cm\,s^{-1}}$ and a column density of $N = 10^{22}\,{\rm cm}^{-2}$. We shall take the total ionizing luminosity of the central engine, $L$, to be $4 \times 10^{43}\,{\rm erg\,s^{-1}}$.

### 3.1 Assumptions

Consider an idealized geometry for the warm absorbing material. Suppose the warm material is distributed in a shell (or blob) with an outer radius $R$ from the central engine and a line-of-sight thickness of $\Delta R$. Assume the material to have a constant density $n$. Thus, the column density is $N = n\Delta R$. Furthermore, we assume that the dominant source of ionizing radiation is the central engine of the AGN and that the warm material is in photoionization equilibrium with this radiation field.

Given the highly variable nature of the observed central ionizing flux, the assumption of photoionization equilibrium requires justification. The recombination time-scales for the ions which dominate the cooling of the gas (mainly highly ionized oxygen and neon) are approximately

$$t_{\rm rec} \sim 3 \times 10^4\, Z^{-2} \left( T_5^{1/2} n_9^{-1} \right)\,{\rm s} \qquad (2)$$

where $Z$ is the atomic number of the ion, $T_e = 10^5 T_5$ K is the electron temperature and $n_e = 10^9 n_9\,{\rm cm}^{-3}$ is the electron density. This uses the approximate form for the recombination coefficients given in Allen (1973), but is in rough agreement with more carefully calculated recombination coefficients (Shull & Van Steenberg 1982). For highly ionized oxygen and neon in the warm state ($T_5 \sim 1$), this evaluates to give a recombination time-scale of a few$\times 10^2 n_9^{-1}$ s. The photoionization time-scale will be significantly less. This is the typical time-scale on which the ionization state of the material responds to changes in the ionizing flux. Unless $n_9 << 1$, this time-scale is shorter than the typical variability time-scale of the primary ionizing continuum in Seyfert galaxies. Photoionization equilibrium would then be expected to apply. However, if $n_9 << 1$ the variability time-scale of the source would become less than the recombination time-scale and photoionization equilibrium would not apply. Under such circumstances the ionization state of the gas would depend on the history of the primary flux variations.

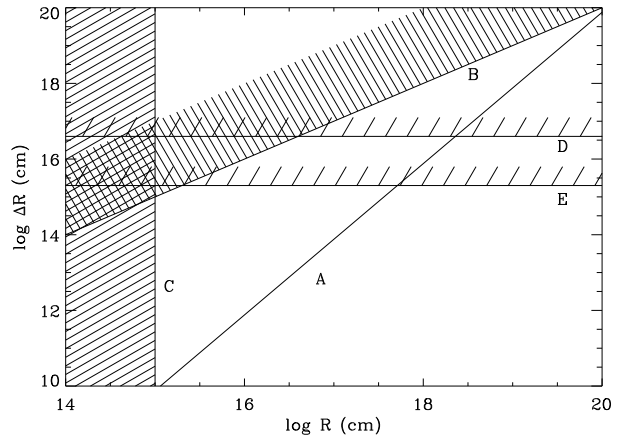

**Figure 2.** Constraints on the $R, \Delta R$ plane as derived in Section 3. Disallowed regions are shaded. Line A corresponds to $\xi = 30\,{\rm erg\,cm\,s^{-1}}$ and $N_{\rm W} = 10^{22}\,{\rm cm}^{-2}$, the approximate parameters of the warm absorbing gas inferred from *ASCA* observations. Within the one-zone photoionization model, the warm gas would lie on this line. Line B is the geometrical constraint that $R > \Delta R$. Line C results from imposing that the virial velocity is less than $0.03\,c$ ($M_7 \sim 1$ assumed in the positioning of this line). Line D gives the constraint that the recombination time-scale is less than 3 weeks. Similarly, line E corresponds to the constraint that drifting inhomogeneities of size $\Delta R$ produce variability in less than 3 weeks.

### 3.2 Basic constraints

Given the assumptions stated, we will now discuss constraints on the $R, \Delta R$ plane. Fig. 2 is a graphical representation of the constraints.

The definition of $\xi$ combined with $N = n\Delta R$ leads to

$$\Delta R = \frac{NR^2 \xi}{L}. \qquad (3)$$

Thus, the warm absorber must lie on this curve within the $R, \Delta R$ plane (line A in Fig. 2). Also, purely geometrical considerations give $R > \Delta R$ (line B in Fig. 2). Together, these conditions give

$$R < \frac{L}{N\xi}. \qquad (4)$$

Using the measured parameters for MCG−6−30−15, this evaluates to give $R < 10^{20}$ cm.

The *ASCA* data require the absorbing material to have radial velocities less than $0.03\,c$, where $c$ is the speed of light. If we impose that the virial velocity of the warm absorbing gas is less than $0.01 v_{0.01} c$, we derive a lower limit to the radius of the warm absorbing material of $R > 10^{16} M_7 v_{0.01}^{-2}$ cm (line C in Fig. 2), where $M = 10^7 M_7\,{\rm M}_\odot$ is the mass of the central compact body. This limit can be relaxed if material in the line of sight to the central engine has a radial velocity much less than the virial velocity. This would be the case for material orbiting the central body in near-circular paths. However, material could not remain on such paths without encountering the accretion disc which is thought to surround the central compact body.

### 3.3   Constraints from variability

*ASCA* observations of MCG−6−30−15 revealed a significant change in the warm absorber between two observations which were carried out 3 weeks apart (Fabian et al. 1994). Simple one-zone photoionization models (produced with CLOUDY) show that the inferred column density increases from $10^{21.8\pm0.1}$ cm$^{-2}$ to $10^{22.13\pm0.02}$ cm$^{-2}$. The inferred ionization parameter also increases slightly from $39^{+7.5}_{-5}$ erg cm s$^{-1}$ to $44.7\pm3$ erg cm s$^{-1}$. This variability can be used to impose stricter (but model dependent) upper limits on $R$.

Suppose the variability is due to warm material responding to changes in the primary ionizing continuum. The inferred recombination time-scale would then have to be less than (or of the order of) 3 weeks. Such a condition leads to a lower limit on the density of the gas of $n > 2.5 \times 10^6$ cm$^{-3}$. A column density of $N = 10^{22}$ cm$^{-2}$ would then give $\Delta R < 4 \times 10^{16}$ cm (line D in Fig. 2) which corresponds to $R < 2 \times 10^{18}$ cm. However, there is mounting evidence that this simple picture for the variability of the warm absorber is invalid. First, in MCG−6−30−15 the ionization parameter is seen to be higher at a time when the ionizing flux is lower (in contrast to the simple photoionization model). Unfortunately, the behaviour of the ionizing flux over the whole 3-week period is unknown; the ionization state of the warm material may have been responding to an unobserved flare between the two *ASCA* observations. Secondly, *ASCA* observations of MR2251−178 show it to have a warm absorber with an ionization parameter that remains essentially constant despite large changes in the 2−10 keV X-ray flux (Kii 1994, private communication). The resolution of this issue will require detailed, time-resolved spectral analysis of the *ASCA* data.

Tangentially moving inhomogeneities in the warm absorbing material can also lead to observable variability. Assuming the primary source to be a point source, and the inhomogeneities to have a characteristic size $\sim \Delta R$ and tangential drift velocity $v$, the variability time-scale is given by

$$t \sim \frac{\Delta R}{v}. \qquad (5)$$

The imposition that $t$ be less than 3 weeks and that $v$ is less than $0.03c$ gives $\Delta R < 2 \times 10^{15}$ cm (line E in Fig. 2), corresponding to $R < 5 \times 10^{17}$ cm. This constraint is weakened or removed if the spatial extension of the primary X-ray source has a size comparable to, or greater than, the characteristic size scale of the inhomogeneities in the warm absorbing material.

Thus we conclude that the warm absorber lies between $10^{15}$ cm and $10^{18}$ cm. This suggests that the warm absorber exists at radii coincident with, or just outside, the BLR. Our results are in accord with Mathur et al. (1994) who find absorption features within the broad UV lines from material that they identify with the warm absorber. Before discussing particular physical models for the origin of this warm material (Section 5), we first examine the thermal stability of this material.

## 4   STABILITY OF THE WARM ABSORBER

Material in the environment of the central engine is prone to thermal instabilities if its temperature lies between $\sim 10^4$ K and the Compton temperature of $10^7$–$10^8$ K (McCray 1979; Krolik, McKee & Tarter 1981; Guilbert, Fabian & McCray 1983). However, the warm absorber appears to represent a significant amount of material in an intermediately ionized state at $10^5$ K. Thus it is interesting to carry out a detailed study of its thermal stability.

Here we address the stability of the warm material under the constraint of isobaric conditions. We use CLOUDY to examine the thermal and photoionization equilibrium in a geometrically thick, optically thin spherical shell of material around a central point source of ionizing flux (with a fixed ionizing luminosity). The shell is constrained to have a constant density (fixed to a given value). Such models allow us to determine the temperature, $T$, as a function of the radius from the ionizing source, $r$. The ionization parameter, $\xi$, can be trivially calculated from its defining equation given that the shell is strictly optically thin. Given $\xi(r)$ and $T(r)$ we can construct the curve on the $T$, $\xi/T$ plane corresponding to thermal equilibrium.

The form of this curve gives information on the thermal stability of the material. Isobaric perturbations have constant $\xi/T$ and thus correspond to vertical displacements on the $T$, $\xi/T$ plane. Consider material in thermal equilibrium. If an isobaric increase in $T$ leads to cooling dominating over heating, then the equilibrium will be stable. However, if such a temperature increase leads to heating dominating over cooling, then runaway heating occurs and the equilibrium is unstable.

Note that the geometrically thick, optically thin shell is not a physical model; it is merely a construction that proves to be convenient in the analysis of the local thermal stability of the material.

### 4.1   Results

Fig. 3 shows such a curve computed for an ionizing continuum consisting of a powerlaw with photon index $\Gamma = 1.8$ extending from 13.6 eV to 40 keV. The density was fixed at $10^9$ cm$^{-3}$ and solar abundances were assumed. The relative dominances of heating and cooling in various regions of the diagram are indicated. It can be seen from Fig. 3 that *parts of the curve that have negative gradient and are associated with a multi-valued regime correspond to thermally unstable equilibria*.

Using the warm absorber parameters obtained from the *ASCA* observations of MCG−6−30−15, it is seen that the warm absorber appears to exist in a small region of stability within an otherwise unstable regime. The region of parameter space for which the warm state is stable is extremely restricted ($-3.48 < \log(\xi/T) < -3.43$ for the case shown in Fig. 3). However, observed column densities and covering fractions suggest that a significant amount of material is in this state. The question arises as to how the material could get into such a state. These issues are addressed in Section 5.

In principle, the exact form of the stability curve depends on the shape of the primary continuum, the assumed density, and the assumed abundances of the material. Thus,

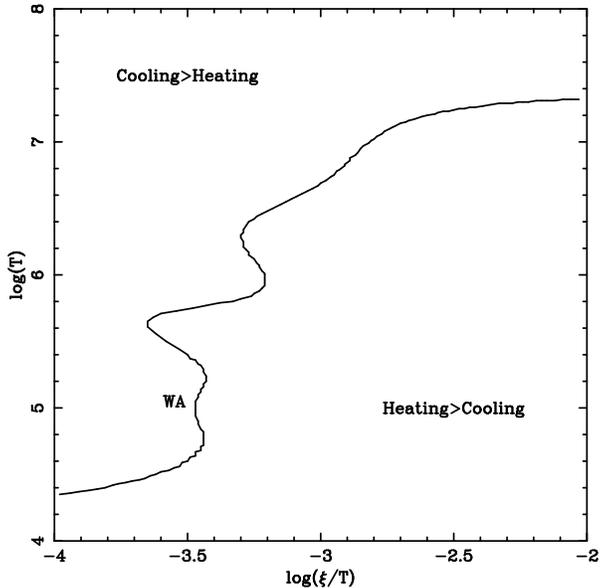

**Figure 3.** Equilibrium gas temperature $T$ as a function of $\xi/T$. The assumed ionizing continuum consists of a single powerlaw with photon index $\Gamma = 1.8$ extending from 13.6 eV to 40 keV. There are small ranges of $\xi/T$ for which $T$ is multi-valued, leading to the possibility of multiple phases in pressure balance. The warm absorber ($T \sim 10^5$ K) is seen to fall in a small stable region within such a multi-phase regime. Regions of the plane where cooling exceeds heating and vice versa are indicated.

the robustness of the stability curve to various changes was examined.

### 4.2 Effect of ionizing continuum shape

First, the effect of changing the photon index of the primary source was investigated. Fig. 4 shows the stability curve for $\Gamma = 1.3$, $\Gamma = 1.8$ and $\Gamma = 2.5$ with fixed low- and high-energy cut-offs at 13.6 eV and 40 keV respectively. For a very flat primary spectrum ($\Gamma < 1.5$) no stable warm state exists. On the other hand, steep primary spectra ($\Gamma > 3$) result in a lowering of the Compton temperature and a stabilization of the material for all $\xi$. The effect of changing the high-energy cut-off was also investigated. The major effect of increasing the value of the high-energy cut-off is to increase the Compton temperature of the gas. Such a change has little effect on the stability of material at temperatures below $\sim 10^6$ K.

The effect of a soft X-ray excess was examined. This was modelled as a blackbody component with a temperature of 0.13 keV. Fig. 5 shows the resulting stability curve for $\Gamma = 1.8$ with varying blackbody luminosities. Moderate soft excesses only change the shape of the curve for $T < 10^5$ K. The trend is for the soft excess to stabilize the cold/warm material. The properties of the hot material ($T > 10^5$ K) are little affected by the soft excess unless it is very strong (e.g. see Fabian et al. 1986), in which case the effect is to lower the Compton temperature.

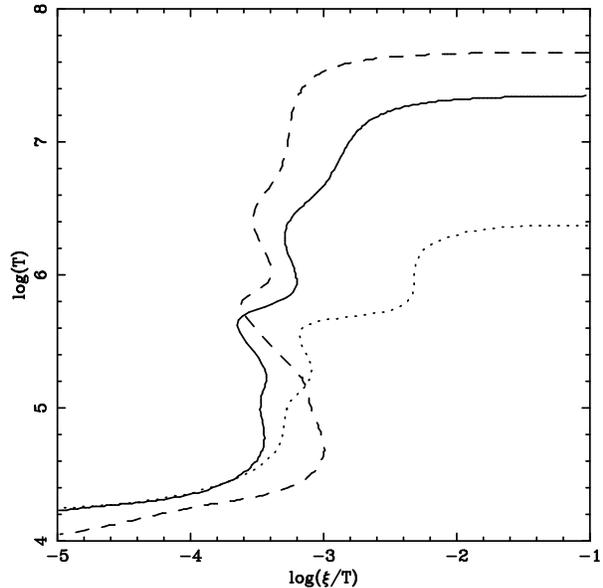

**Figure 4.** Equilibrium gas temperature $T$ as a function of $\xi/T$ for photon indices of $\Gamma = 1.3$ (dashed curve), $\Gamma = 1.8$ (solid curve) and $\Gamma = 2.5$ (dotted curve). In all cases the powerlaw extends from 13.6 eV to 40 keV.

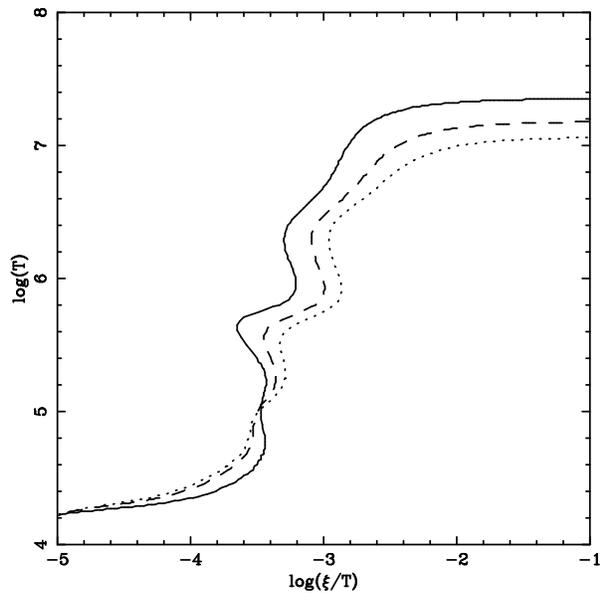

**Figure 5.** Equilibrium gas temperature $T$ as a function of $\xi/T$ for a photon index of $\Gamma = 1.8$ and various soft excess luminosities. The soft excesses are modelled by a power-law component with a temperature of 0.13 keV. The solid line is the curve of Fig. 3 with no soft excess. The dashed and dotted lines are the results for a soft excess with 50 per cent and 100 per cent of the power-law luminosity respectively. In all cases the powerlaw extends from 13.6 eV to 40 keV.

### 4.3 Density effects

Next the sensitivity to the assumed density was examined. The curve on the $T$, $\xi/T$ plane was computed for various densities between $0.1\,\mathrm{cm^{-3}}$ and $10^{13}\,\mathrm{cm^{-3}}$. For densities between $10\,\mathrm{cm^{-3}}$ and $10^{11}\,\mathrm{cm^{-3}}$ the form of the curve is very insensitive to the assumed density. This is to be expected: consider material with such densities. If it is cold/warm ($T < 10^6$ K) the thermal equilibrium is determined by the balance between photoelectric heating and cooling due to atomic lines (see Netzer 1990 for a discussion of the physical processes). The density will then enter only via the combination $\xi$. If the material is hot ($T > 10^6$ K) the thermal equilibrium is determined mainly by the balance of Compton heating and Compton cooling. Thus it will be independent of density.

For densities greater than $10^{11}\,\mathrm{cm^{-3}}$, three-body recombination becomes important whereas, for densities below $10\,\mathrm{cm^{-3}}$, two-body cooling processes become inefficient. These effects produce a more complex dependence on density thereby affecting the form of this curve. However, the arguments of Section 3 suggest that these extreme densities are not relevant to the observed warm absorbers.

### 4.4 Finite optical depth effects

Observed warm absorber column densities are $\sim 10^{22}\,\mathrm{cm^{-2}}$. Thus the optical depths near the dominant absorption edges (i.e. the O VII and O VIII edges) are not negligible. As the ionizing continuum passes through the warm material, the absorption of flux near the O VII and O VIII edges could alter the *shape* of the stability curve of material at the outer edge of the shell. To assess the importance of this effect, CLOUDY was used to compute the transmitted spectrum which emerges from the warm absorbing material (given the parameters obtained for MCG$-$6$-$30$-$15). The stability curve for this spectrum was then computed using the optically thin, geometrically thick shell (as above). It is found that warm absorption does little to change the shape of the stability curve.

## 5 PHYSICAL MODELS OF WARM ABSORBERS

Models for the environment within the nuclei of Seyfert galaxies must explain the presence of significant quantities of warm material. More general AGN models must also explain the general absence of observed warm absorption features in radio-loud objects and powerful radio-quiet QSOs. The purpose of this section is to discuss several models for the origin of the warm absorber.

The arguments of Section 3 suggest that the warm material lies between $10^{15}$ cm and $10^{18}$ cm from the central engine (using the parameters for MCG$-$6$-$30$-$15). This places it roughly coincident with, or just outside, the BLR. Therefore, we shall mainly discuss the warm absorber in relation to various BLR models. Radiatively driven outflows will also be discussed as a possible source for the observed warm absorption.

### 5.1 Pressure-confined BLR

The simplest model for the BLR is the two-phase model (McCray 1979; Krolik et al. 1981). If the thermal instability of Section 4 can operate over a wide range of $\xi/T$, there will be a range of pressure and mean density in which cold ($T \sim 10^4$ K) clouds exist in pressure equilibrium with a hot phase at the Compton temperature ($\sim 10^7$–$10^8$ K). The cold clouds comprise the broad emission line clouds. The cold clouds acquire velocities characteristic of the virial velocity, thereby leading to (Doppler) broadening of the emission lines as required by observation.

Section 4 shows that there is a small range of $\xi/T$ for which a three-phase medium may occur. This would consist of cold (broad-line) clouds in pressure equilibrium with a hot intercloud medium and a warm intermediate phase. If the covering fraction of the cold phase is small, it is plausible that only the warm phase would have sufficient covering fraction and X-ray opacity to be commonly seen in absorption. As a model of the warm absorber, this suffers from problems. It requires fine tuning of $\xi/T$ (and therefore the pressure) to produce the three-phase medium. This model also fails to explain the absence of warm absorbers in powerful QSOs and radio-loud quasars which seem to display ordinary BLRs.

More seriously, even the simple two-phase BLR model has serious flaws. First, in low-luminosity systems (such as Seyfert galaxies) the cooling time-scale of the intercloud medium is longer than the dynamical time-scale of the BLR. Thus the assumption of thermal equilibrium need not be valid. This was noted by Krolik et al. (1981, hereafter KMT). Secondly, the formation of broad-line clouds is not explained within this model. Compton cooling time-scales are too long to allow perturbations of the hot phase to condense into clouds. Also, self-gravity of the clouds is thought to be negligible, thereby ruling out gravitational instabilities as a formation mechanism. In this sense, the simple two-phase model is incomplete. Thirdly, Fabian et al. (1986) showed that a strong soft X-ray/EUV excess will reduce the Compton temperature to $\sim 10^6$ K *and* severely restrict the range of $\xi/T$ over which the thermal instability can occur. This result has several consequences.

(i) The intercloud medium would have to be optically thick if it were still to pressure-confine the broad-line clouds. An optically thick intercloud medium is inconsistent with the observed rapid variability of many broad-line AGN.

(ii) The ionization parameter of the broad-line clouds can be determined from the observed line ratios. These observed values of $\xi$ span a larger range than that over which the thermal instability operates. Such discrepancies can be reconciled if some physical process (such as cloud evaporation or partial magnetic confinement) increases the pressure of the clouds above that of the intercloud medium (Kallman & Mushotzky 1985).

(iii) Broad-line clouds are inferred to have velocities of up to $\sim 10^4\,\mathrm{km\,s^{-1}}$. Such motion through a stationary intercloud medium at $T \sim 10^6$ K would lead to rapid fragmentation of the clouds into optically thin filaments, contrary to the observed line ratios. This difficulty would be alleviated if the broad-line clouds were advected in some large-scale, high-velocity flow of the intercloud medium.

Finally, recent results from reverberation mapping (Pe-terson 1993) suggest that the BLR is extended with an inner radius at least 10 times smaller than the outer radius. The two-phase model of the BLR would predict a relatively thin shell-like BLR.

A full description of the BLR clearly has to be dynamic in nature. We now discuss some dynamic models for the BLR.

### 5.2 Shock-formed BLR

Perry & Dyson (1985) have proposed a model in which the broad-line clouds are continuously created by radiative shocks. A supersonic flow of hot material (e.g. a wind from the central source) encounters a number of large obstacles, such as supernova ejecta or winds from groups of massive stars. The resulting shocks compress and heat the gas, thereby dramatically reducing $\xi/T$ in the post-shocked gas. Provided the shocks are spatially large, the post-shocked gas can cool isobarically via inverse Compton scattering of the radiation field from the central engine. This gas cools to almost $10^4$ K and fragments to form broad-line clouds. After the clouds leave the environment of the shock, they are accelerated to the local flow speed (thereby giving the broad-line velocities) and eventually evaporate. No cloud confinement is necessary due to their continuous production.

Within the context of this model, the warm absorber could be identified with material that has been 'boiled' from the clouds and is being heated to the Compton temperature. Further investigation is required to examine whether the necessary column densities, covering fractions and ionization states can be achieved. It may, for example, be difficult to inflate broad-line clouds with a covering fraction of a few per cent and a column density of $\sim 10^{22}$ cm$^{-2}$ to produce a warm absorber covering fraction of $\sim 50$ per cent and a similar column density.

A crucial component of this model is the population of large obstacles which create the spatially extended shocks in the hot gas. Reverberation mapping of the BLRs of Seyfert galaxies suggest it to be surprising small, with spatial extents of only $\sim 20$ light days or so. There would have to be a large number of obstacles within this radius to produce the observed, fairly regular BLR. This might pose problems for the model.

### 5.3 Turbulent BLR

Mechanical heating of the hot intercloud medium has been given little attention in recent BLR models. If mechanical heating could raise the temperature of the intercloud medium to $\sim 10^8$ K or more, the problems associated with a pressure-confined BLR (see Section 5.1) would be alleviated. Here we examine turbulent heating of the intercloud medium.

Suppose the hot intercloud medium (HIM) has a temperature of $T \sim 10^8$ K and a density of $n \sim 10^6$ cm$^{-3}$ so that approximate pressure balance with the broad-line clouds is achieved. The thermal energy density in the electrons is

$$E_{\rm HIM} = \frac{3}{2} n k_{\rm B} T \sim 10^{-2} \; {\rm erg \, cm^{-3}}. \tag{6}$$

The Compton cooling time-scale is given by

$$t_{\rm Comp} \sim \frac{m_{\rm e} c^2 R^2}{L \sigma_{\rm T}} \tag{7}$$

where $m_{\rm e}$ is the rest mass of an electron, $\sigma_{\rm T}$ is the Thomson cross section and $R = 10^{16} R_{16}$ cm is the distance from the central source which is assumed to have an isotropic luminosity of $L = 10^{43} L_{43}$ erg s$^{-1}$. This evaluates to give $t_{\rm Comp} \sim 10^8 \, R_{16}^2 L_{43}^{-1}$ s. The Compton cooling rate per unit volume is

$$W_{\rm Comp} \sim \frac{E_{\rm HIM}}{t_{\rm Comp}} \sim 10^{-10} \, R_{16}^{-2} L_{43} \; {\rm erg \, cm^{-3} \, s^{-1}}. \tag{8}$$

If additional heat sources can exceed this power per unit volume, they will dominate over Compton cooling and hold the intercloud medium significantly above the Compton temperature. We now estimate the turbulent heating and show that it could dominate Compton cooling in the BLRs of Seyfert galaxies.

Radiation pressure on broad-line clouds will induce motion of the clouds with respect to the intercloud medium. The terminal velocity $v_{\rm t}$ is given by $P_{\rm rad} \approx \rho v_{\rm t}^2$ where $P_{\rm rad}$ is the radiation pressure on the clouds and $\rho$ is the mass density of the intercloud medium. The radiation field thereby does work on the broad-line clouds. The rate of doing work on the clouds is $\sim P_{\rm rad} v_{\rm t}$ per unit illuminated area. Therefore, the rate of doing work per unit volume is

$$W \sim \frac{\mathcal{N} \mathcal{A} P_{\rm rad}^{3/2}}{V \rho^{1/2}} \tag{9}$$

where $\mathcal{N} = 10^6 \mathcal{N}_6$ is the number of broad-line clouds (with $\mathcal{N}_6 > 1$), $V$ is the volume of the broad-line region and $\mathcal{A}$ is the projected area of a single broad-line cloud. If we define $r = 10^{12} r_{12}$ cm to be the typical radius of a broad-line cloud ($r_{12} \sim 1$) and use $P_{\rm rad} = L/4\pi R^2 c$, we deduce that

$$W \sim 10^{-10} \mathcal{N}_6 r_{12}^2 R_{16}^{-6} L_{43}^{3/2} \; {\rm erg \, cm^{-3} \, s^{-1}}. \tag{10}$$

Expressed in terms of the covering fraction, $f_{\rm A}$, of the broad-line clouds, this becomes

$$W \sim 10^{-10} f_{\rm A} R_{16}^{-4} L_{43}^{3/2} \; {\rm erg \, cm \, s^{-1}}. \tag{11}$$

It is reasonable to suppose that this energy is transferred into large-scale turbulent motions within the intercloud medium. Turbulent cascades would lead to thermalization of this energy in the intercloud medium. Such a mechanism would provide an additional heat source for the intercloud medium. The ratio of the turbulent heating rate to the Compton cooling rate is

$$\mathcal{R} = \frac{W}{W_{\rm Comp}} \sim \mathcal{N}_6 r_{12}^2 R_{16}^{-4} L_{43}^{1/2}. \tag{12}$$

Turbulent heating will dominate Compton cooling if $\mathcal{R} > 1$. Such a condition may well be satisfied in Seyfert galaxies. The characteristic size $R$ of the BLR is thought to be proportional to $L^{1/2}$ (i.e. the characteristic BLR ionization parameter is similar in objects with very different luminosities). Assuming that cloud size, $r$, is independent of the total luminosity of the object, this gives $\mathcal{R} \propto \mathcal{N} R^{-3}$. Thus, turbulent heating can exceed Compton heating (i.e. $\mathcal{R} > 1$) in more luminous AGN only if the number of broad-line clouds, $\mathcal{N}$, increases at least as fast as $R^3$.

The radiation pressure on the warm absorbing material is only a small fraction of the radiation pressure on the broad-line clouds (of the order of a few per cent). However, the warm material has a much greater covering fraction. Thus, the action of the radiative forces on the warm material might also be important for turbulent heating of the intercloud medium.

Turbulence in the hot intercloud medium would disrupt the surfaces of the broad-line clouds. The cold and hot material would then mix to form a turbulent mixing layer. Begelman & Fabian (1990) use conservation of energy and momentum to deduce that a steady-state mixing layer is characterized by a temperature

$$T_{\rm ml} = \eta (T_{\rm c} T_{\rm h})^{1/2} \qquad (13)$$

where $T_{\rm c}$ and $T_{\rm h}$ are the temperatures of the cold and hot phases respectively and $\eta \sim 1$. For $T_{\rm c} \sim 10^4$ K and $T_{\rm h} \sim 10^8$ K this gives $T_{\rm ml} \sim 10^6$ K. Mixed gas could shear from the surfaces of the broad-line clouds to produce optically thin filaments. Such filaments could be identified with the warm absorber.

Mixing layers will be strong emitters of UV radiation. An estimate for the resultant UV flux can be obtained by considering the thermal energy of the intercloud medium that is 'swept up' by the broad-line clouds each unit of time. If we assume that this energy is completely converted to UV flux, we obtain an approximate upper limit for the total UV flux due to mixing layers. This flux proves to be negligible compared with the UV flux of the central engine and has no effect on the ionization structure of the broad-line clouds.

Mixing layers are complex, non-equilibrium systems and are examined in detail by Slavin, Shull & Begelman (1993) for the case of the interstellar medium. They contain cooling components as well as components undergoing heating. Self-photoionization must also be considered. The AGN environment is very different from the interstellar medium: the pressures are much greater, there are high-speed bulk flows and there is a luminous source of hard radiation. Thus, more investigation of mixing layers in AGN is required to assess the observational consequences.

### 5.4 Outflow models

Recent *ASCA* observations of the warm absorber in NGC 4051 hint that the absorption edges of O VII, O VIII, Ne IX and Ne X may be blueshifted by $\sim 3$ per cent (Mihara et al. 1994). If verified, this suggests an outflow of the material at $\sim 10000$ km s$^{-1}$. Since radiation pressure on the material can be comparable with the gravitational attraction of the central compact body, the possibility of a radiatively driven wind arises. Here we sketch some simple ideas relating to warm absorbers as radiatively driven outflows.

It is instructive to calculate the critical ionizing luminosity for which the radiative force on the warm absorber balances the gravitational attraction of the central compact body. For the purposes of obtaining a crude estimate, suppose that all of the oxygen is in the form of O VII and that the dominant opacity of the material is K-shell photoionization of this ion. Take the source of the ionizing continuum to be an isotropic emitter of a power-law spectrum with photon index $\Gamma = 2$ between $\nu_{\rm min}$ and $\nu_{\rm max}$. Let this source be situated at $R = 0$. Thus

$$L(\nu) = \frac{L}{\Lambda} \nu^{-1} \qquad (14)$$

where $L$ is the total ionizing luminosity and

$$\Lambda = \ln \left( \frac{\nu_{\rm max}}{\nu_{\rm min}} \right). \qquad (15)$$

The radiative force per O VII ion is given by

$$F_{\rm rad} = \int_0^\infty \frac{L(\nu)}{4\pi c R^2} \sigma(\nu) {\rm d}\nu \qquad (16)$$

where $\sigma(\nu)$ is the cross-section for K-shell photoionization of O VII. To a fair approximation this can be taken to be proportional to $\nu^{-3}$ for $\nu > \nu_{\rm th}$, the threshold frequency where $\sigma = \sigma_{\rm th}$ (and zero below $\nu_{\rm th}$). Thus we carry out the integration in the previous expression to give

$$F_{\rm rad} = \frac{L \sigma_{\rm th}}{12 \pi c \Lambda R^2}. \qquad (17)$$

Now suppose the gravitational potential of the region is dominated by a mass $M$ at $R = 0$. The gravitational force on the gas per O VII ion is

$$F_{\rm grav} = \frac{G M m}{R^2} \qquad (18)$$

where $m$ is the mass of gas per O VII ion. Thus, under the simplifying assumptions stated, the ratio of radiative force (radially outwards) to gravitational force (radially inwards), $\Phi$, is given by

$$\Phi = \frac{1}{3\Lambda} \left( \frac{L}{L_{\rm E}} \right) \left( \frac{m_{\rm p}}{m} \right) \left( \frac{\sigma_{\rm th}}{\sigma_{\rm T}} \right) \qquad (19)$$

where $L_{\rm E}$ is the Eddington luminosity, $m_{\rm p}$ is the proton mass and $\sigma_{\rm T}$ is the Thomson cross-section. Using $\sigma_{\rm th} = 2.75 \times 10^{-19}$ cm$^2$ and approximate solar abundances (so that $m \sim 10^3 m_{\rm p}$), we obtain

$$\Phi \sim 20 \left( \frac{L}{L_{\rm E}} \right) \qquad (20)$$

Thus the critical luminosity for which $\Phi = 1$ is $L_{\rm crit} \sim 0.05 L_{\rm E}$. This is a plausible value for many Seyfert galaxies. So, for $L = L_{\rm crit}$ the steady-state radiatively driven wind consists of material in the warm state. Outflowing material that is fully ionized (with correspondingly lower opacity) would be gravitationally decelerated and thus compressed until it recombines to give warm material. Similarly, neutral material (with correspondingly higher opacity) would be radiatively accelerated and rarefied until it partially ionizes to give warm material. Such an argument could explain the unusual ionization state of the warm absorber.

Continuing with this simple model, suppose $L \sim L_{\rm crit}$ and that this produces a homogeneous outflow of warm material with constant velocity $v$ beyond some radius $r_{\rm in}$ within a solid angle $\Omega$. Neglecting the region $r < r_{\rm in}$, the optical depth of the flow is given by

$$\tau(\nu) \approx \sigma(\nu) N({\rm O\ VII}) \qquad (21)$$

where $N({\rm O\ VII})$ is the column density of O VII ions. If the flow becomes optically thick over a significant range of frequencies, the radiative force will significantly increase and the steady state will be broken. Thus, we shall impose that $\tau(\nu) < 1$ for all $\nu$. In particular, $\tau(\nu_{\rm th}) < 1$. This yields an upper limit on the column density of O VII ions of $N({\rm O\ VII}) < 4 \times 10^{18}$ cm$^{-2}$. This translates into an equivalent hydrogen column density of $N_{\rm W} < 4 \times 10^{21}$ cm$^{-2}$. Observationally, warm absorbers are seen with $N_{\rm W} \sim 10^{22}$ cm$^{-2}$.

Thus there is crude agreement between the observed column density and the maximum column density allowed by this model under the constraint of optical thinness. Assuming no mass injection in the region $r > r_{\rm in}$, conservation of mass gives $n(r) \propto r^{-2}$. The mass outflow rate (which is given by $r^2 \Omega v \rho$) is then

$$\dot{M} \sim 3 \times 10^{-4} \Omega \left( r_{16}^{\rm in} N_{22}^{\rm W} v_3 \right) \, {\rm M}_\odot \, {\rm yr}^{-1} \qquad (22)$$

where $r_{\rm in} = 10^{16} r_{16}^{\rm in}$ cm, $N_{\rm W} = 10^{22} N_{22}^{\rm W}$ cm$^{-2}$ and $v = 10^3 v_3$ km s$^{-1}$ The kinetic luminosity is

$$L_{\rm K} \sim 10^{38} \Omega \left( r_{16}^{\rm in} N_{22}^{\rm W} v_3^3 \right) \, {\rm erg \, s}^{-1}. \qquad (23)$$

Thus, even quite high-velocity outflows would not dominate the energetics of the source.

Now suppose $L < L_{\rm crit}$. Our arguments suggest that a steady-state outflow of cold (partly neutral) material might form. This material would have a sufficiently high opacity for the radiative force to balance gravitation. Similarly, if $L > L_{\rm crit}$ an outflow of more highly ionized material may be created. This scenario could account for the observed anticorrelation between the AGN luminosity and intrinsic column density for cold absorption (Reichert et al. 1985; Turner & Pounds 1989). It could also explain the absence of observed warm absorbers in powerful QSOs (which are thought to be radiating close to the Eddington limit).

Clearly, a detailed study of such radiatively driven winds would be complicated. Such a treatment would have to include realistic opacities, the possibility of an inhomogeneous flow (e.g. break-up of the flow into clouds via the thermal instability of the previous sections), anisotropy of the primary ionizing continuum and additional heating and cooling mechanisms (e.g. shocks, turbulence and adiabatic cooling).

## 6 SUMMARY

Absorption features due to partially ionized, warm ($\sim 10^5$ K) gas are common in the X-ray spectra of Seyfert 1 galaxies and NELGs. The *ASCA* SIS allows such features to be accurately identified and measured. The dominant features in the *ASCA* energy range are absorption K-edges of O VII and O VIII. Ne IX and Ne X edges are also observable in some sources. One-zone photoionization models allow the column density and ionization parameter $\xi$ of the warm material to be estimated (assuming that the material is in photoionization equilibrium). The covering fraction of this material within Seyfert 1 nuclei and NELGs must be large for this phenomenon to be so commonly observed.

Constraints on the location of the warm absorbing gas are derived using the simplifying assumptions that the material is of uniform density and occupies a region with line-of-sight depth (to the central engine) $\Delta R$ and outer radius $R$ from the central engine. Thermal equilibrium and photoionization equilibrium with the radiation field from the central engine are also assumed. We conclude that the warm material must lie between $10^{15}$ cm and $10^{18}$ cm from the central engine. This implies that the warm material is spatially coincident with, or just outside, the BLR.

Intermediately ionized plasma is notoriously unstable to temperature fluctuations. Thus, the stability of the warm absorbing material requires investigation. We have done this by constructing the equilibrium curve on the $T$, $\xi/T$ plane. Parts of this curve that have negative gradient *and* are associated with a multi-valued regime correspond to thermally unstable equilibria. Use of a standard ionizing continuum leads to the conclusion that a stable warm state only exists for a remarkable small range of $\xi$. A steep primary spectrum or a very strong soft excess stabilizes all material cooler than $10^5$ K. These results are insensitive to reasonable changes in the assumed density.

Given the small range of $\xi$ for which the warm state is stable, we must explain how such a large amount of material attains such a state. The absence of observed warm absorbers in most radio-loud objects and powerful radio-quiet QSOs must also be explained. In an attempt to address these issues, several models of the warm absorber are discussed. In particular, possible relationships between the warm absorber and the BLR are examined. Within the context of the pressure-confined two-phase BLR of KMT, a three-phase medium may form in which cold broad-line clouds exist in pressure equilibrium with a hot intercloud medium and an intermediate warm phase. It is plausible that only the warm phase would have a sufficiently high covering fraction and X-ray opacity to be seen commonly in absorption. However, the simple two-phase model is incomplete (in the sense that it neglects dynamical effects and does not explain the formation of broad-line clouds) and suffers from inherent difficulties such as an optically thick intercloud medium and rapid fragmentation of the broad-line clouds into optically thin filaments (both of which are contrary to observations).

Radiation pressure acting on broad-line clouds can be significant. The work done on the clouds by the radiation field can drive turbulent motions in the intercloud medium. This energy undergoes a cascade to smaller and smaller scales until it is thermalized. Such a mechanism provides a mechanical heat source for the hot intercloud medium. We suggest that this can dominate Compton heating of the intercloud medium in the BLRs of Seyfert galaxies. Turbulence in the intercloud medium leads to the formation of turbulent mixing layers at the surface of the broad-line clouds. Such mixing layers may give rise to optically thin filaments of warm material, thereby providing an explanation for the observed warm absorption features.

Finally, a simple model for a steady-state radiatively driven outflow is examined. It is concluded that, for $L \sim 0.05 L_{\rm E}$, an optically thin outflow of warm material could form. Lower luminosity objects would have outflows of colder (more recombined) material whereas higher luminosity objects would have outflows of hotter material (more ionized). This explains the apparent anti-correlation of cold absorbing column density with luminosity. It also explains the absence of warm absorbers in powerful QSOs (assuming these objects to be accreting closer to the Eddington limit than their lower luminosity counterparts).

In a recent paper by Krolik & Kriss (1994), seen by us after the submission of this work, it was emphasized that the absorbing material may not be in thermal equilibrium and that this may have important implications for the resultant spectrum. Non-equilibrium conditions might be expected to be relevant to the shock-formed BLR, turbulent BLR and outflow models described above in which mechanical heating may be important.

The physical state and time variability of this warm ma-terial could be important tools for probing the environment near the central engines of Seyfert galaxies. To date, studies of the BLR have been constrained only by observations of the broad lines (and upper limits on quantities such as the optical depth or X-ray emission of the intercloud medium). If the connection between the warm absorber and BLR is valid, the properties of this warm material could be an important diagnostic of the BLR. Furthermore, the presence or absence of warm absorbers in the various classes of object can be used to assess unified AGN models.

## ACKNOWLEDGMENTS

We wish to thank Niel Brandt, Alastair Edge, Roderick Johnstone, Tsuneo Kii, Paul Nandra and Martin Rees for useful discussions throughout the course of this work. We also wish to thank Gary Ferland for his valuable help in the operation of CLOUDY. ACF thanks the Royal Society for support. CSR thanks PPARC for support.